\documentclass[pra,superscriptaddress]{revtex4}

\usepackage{amsmath}
\usepackage{amsfonts} 
\usepackage{graphicx}
\usepackage{dcolumn}
\usepackage{epsfig}
\usepackage{bm}
\usepackage{array}
\usepackage{hyperref}
\hypersetup{
colorlinks=true,
citecolor=blue,
linkcolor=blue,
urlcolor=blue,
}
\usepackage{xcolor}	 % Color features
\usepackage{graphicx}
\usepackage{amsmath}
\usepackage{comment}
\usepackage{amssymb}
\usepackage{bm} % Bold math
\usepackage{bbm}
\usepackage[english]{babel}
\usepackage{setspace}
\usepackage{wasysym}
\usepackage{adjustbox}
\usepackage{framed}
\usepackage{mdframed}
\usepackage{latexsym}
\usepackage{pdflscape} % for horizontal page
\usepackage{multirow} % some options for tabulars
\usepackage{booktabs} % some options for tabulars\\
\usepackage{calc} % for "\widthof" for measuring the wide of some content
\usepackage{calrsfs} % use $\mathcal{ABC}$ for this fancy calligraphic fonts
\DeclareMathAlphabet{\pazocal}{OMS}{zplm}{m}{n}	% use $\pazocal{ABC}$ for the standard calligraphic fonts

\definecolor{DeepSkyBlue}{RGB}{0,104,139}
\colorlet{MySky}{white!40!blue}
\colorlet{MyViolet}{red!45!blue}
\colorlet{MyBlue}{black!40!blue}
\colorlet{MyRed}{black!40!red}
\colorlet{MyOrange}{red!70!yellow}
\colorlet{MyGreen}{black!60!green}
\colorlet{MyBrown}{black!70!brown}
\colorlet{MyGray}{black!60!white}

\newcommand{\be}{\begin{equation}}
\newcommand{\ee}{\end{equation}}

\newcommand{\beq}{\begin{eqnarray}}
\newcommand{\eeq}{\end{eqnarray}}

\newcommand{\ket}[1]{\ensuremath{|#1\rangle}}

\begin{document}

\title{Revealing the topology of quantum states via Kirkwood-Dirac quasiprobabilities}

\author{Stefano Gherardini}
\affiliation{Istituto Nazionale di Ottica (INO), Consiglio Nazionale delle Ricerche, Largo Enrico Fermi 6, I-50125 Firenze, Italy.}
\affiliation{European Laboratory for Non-linear Spectroscopy (LENS), Università di Firenze, I-50019 Sesto Fiorentino, Italy.}

\author{Luca Lepori}
%\email[Correspondence at: ]{llepori81@gmail.com}
\affiliation{Dipartimento di Scienze Matematiche, Fisiche e Informatiche, Universit\`a  di Parma, Parco Area delle Scienze, 53/A, I-43124 Parma, Italy.}
\affiliation{Gruppo Collegato di Parma, INFN-Sezione Milano-Bicocca, I-43124 Parma, Italy.}
\affiliation{UdR Parma, INSTM, I-43124 Parma, Italy.}

\begin{abstract}
We discuss a theoretical approach to discriminate whether two states of a many-body quantum system belong or not to different topology classes. This approach is based on expressing a strange correlator---a recently established tool for quantum topology discrimination---between the states as a function of Kirkwood-Dirac quasiprobabilities (KDQs). KDQs provide a first-principles representation of two-time quantum correlators. 
{\color{black}The link between strange correlators and KDQs allows to establish that strange correlators are weak values of an observable converting an initial trivial state into a topologically non-trivial one. We thus propose a quantum topology witness that is achievable measuring the prior and subsequent effects on a many-body system of a sudden quench transformation that realizes the transition between trivial and topological phases.} The witness is evaluated on a probe quantum state whose main features are detailed within the paper. {\color{black} Finally, directly exploiting schemes that allows for the complete reconstruction of KDQs, we address an interferometric protocol} for topology discrimination, along with a general discussion of the main lines and challenges towards its implementation.
\end{abstract}

\maketitle

%%%%%%%%%%%%%%%%%%%%%%%%%
\section{Introduction}

Topology and the physics of topological matter have garnered increasing attention over the past four decades, driving substantial theoretical progress and yielding numerous applications in fields ranging from condensed matter to elementary particle physics~\cite{bernevigbook,qi2010,gentini2024,coleman1,coleman2,Nakahara,rothe,volovik2003}.

Two types of topologically ordered states are generally recognized: symmetry-protected topological order, where, as the name itself suggests, topology relies on certain (mostly discrete) global symmetries of the system, and genuine topological order, where no symmetry protects and constrains topology. The first set includes topological insulators and superconductors~\cite{altland1997,ludwig2009,haldanelec} and crystalline generalizations~\cite{fu2011}, while the second set includes fractional quantum Hall states, certain quantum spin liquids, and toric codes~\cite{wen2010,fradkinbook,kitaev2003,laughlinlec}. A further well-known instance of a topologically ordered state is given by Berezinskii–Kosterlitz–Thouless (BKT) phases~\cite{jorge2013,kostlec}.

Topologically trivial matter can generally be described by a {\it local} order parameter (in some coordinate space, such as coordinate or momentum space), typically strongly constrained by both the dimensionality and by the symmetry content of the system. In contrast, topological matter is described by  {\it nonlocal} order parameters, possibly blind to (and thus not affected by) the aforementioned symmetry content. As a result, characterizing topological matter often involves computing complicated integrals over a compact domain, e.g.~Brillouin zones (as for topological insulators and superconductors) or on degenerate ground-state manifolds (as for genuine topological order). These integrals are nothing but suitable reformulations of mathematical topological invariants for the system under analysis. Unfortunately, these quantities are not always straightforward to be related to experimentally-accessible observables.

Due to this general difficulty, many fine-tuned detection strategies have been developed over the years. These strategies include probing edge states via conductivity~\cite{vonk1980,vklitzinglec,foxon1988,konig2007}, ARPES and imaging techniques \cite{hasan2009,chen2009,yazdani2012}, or through interferometric measurements~\cite{interf1,interf2}. More recently, in ultracold atom experiments, Berry-phase topological invariants for Bloch bands have been measured directly, again via interferometry~\cite{duca2015,duca2016}.

Another route to reveal and to probe topological quantum states has been theoretically formulated in the last years, based on introducing the so-called strange correlators~\cite{xu2014,lepori2022}. This concept stimulated an intense investigation, mostly applying it to various systems, and a remarkable consequent literature (see citing works of \cite{xu2014}). Given a pure quantum state $\ket{\Psi}$ of a many-body system defined on a lattice, a strange correlator is defined as
\begin{equation}
\label{matper}
s[\hat{o}]_{{\bf r}, {\bf r^{\prime}}} = \frac{ \langle\Omega| \hat{o}({\bf r}) \hat{o}({\bf r^{\prime}}) |\Psi\rangle }{ \langle\Omega| \Psi \rangle},
\end{equation}
where $\ket{\Omega}$ is another pure quantum state with known trivial topology, such as a fully separable state, and $\hat{o}({\bf r})$ is a many-body operator, with ${\bf r}$ denoting collectively the lattice coordinates. The main difference between strange correlators and ordinary two-point correlation functions in coordinate space is that the involved ket and bra states are different. It has been shown~\cite{xu2014,lepori2022} that, for a suitable choice of the local operators $\hat{o}({\bf r})$, then $s[\hat{o}]_{{\bf r}, {\bf r^{\prime}}}$ scales as $\sim \frac{1}{|{\bf r} - {\bf r}^{\prime}|^a}$, with $a \geq 0$, if $\ket{\Psi}$ has nontrivial topology. Otherwise, if $\ket{\Psi}$ has also trivial topology, then $s[\hat{o}]_{{\bf r}, {\bf r^{\prime}}}$ decays exponentially, as in standard two-point correlations of a massive theory.  More in general, this scaling occurs whenever $\ket{\Omega}$ and $\ket{\Psi}$ belong to different topology classes or host different topologies in the same class.

As anticipated, the algebraic scaling behavior described above holds only for particular system-dependent choices of $\hat{o}({\bf r})$. In \cite{lepori2022}, it has been shown how to identify these choices. Specifically, $\hat{o}({\bf r})$ must be chosen as annihilation operators of bulk excitations that evolve into edge states (in topological phases) when the boundary conditions are smoothly changed from closed (e.g.~periodic) to open (at least along one direction). In physically relevant (including genuine) topological systems~\cite{xu2014,lepori2022}, this structure and the corresponding operators $\hat{o}({\bf r})$ can be often identified. The effectiveness of the strange correlators as witnesses of nontrivial topology has been demonstrated~\cite{lepori2022} for Majorana and AKLT chains, integer quantum Hall states, and Laughlin (Abelian fractional quantum Hall) states, both with closed and periodic boundary conditions.

In this work, we discuss how strange correlators can be expressed in terms of Kirkwood-Dirac quasiprobabilities (KDQs)~\cite{DeBievrePRL2021,LostaglioQuantum2023,DeBievreJMathPhys2023,GherardiniTutorial,hernandez2024Interfero,ArvidssonShukur2024review}, i.e., as two-time quantum correlators built from both $\hat{o}({\bf r})$ and from the pure quantum states $|\Omega\rangle$ and $|\Psi\rangle$. First, this conceptual link is important for interpreting how strange correlators reveal topological matter. This link suggests that their efficacy stems from the noncommutativity of the operator {\color{black}$\hat{O}$ (as denoted below)} that generates the sudden transition between trivial and topological phases, that is from the corresponding pre- and post-quench states. {\color{black}Moreover, from an operational point of view, our analysis allows to interpret the strange correlators as a {\it weak value}~\cite{vaidman1988} of $\hat{O}$. Weak values can serve to carry out quantum state tomography~\cite{DresselRMP2014}, quantum thermodynamics and metrology~\cite{LostaglioPRL2020}, and proofs of contextuality~\cite{PuseyPRL2014}. To our knowledge, this interpretation of strange correlators is novel, and we expect it could open up new research directions within the fields where weak values have found application, even experimentally.} Second, leveraging this theoretical connection {\color{black}between strange correlators and KDQs}, we address a detection strategy based on interferometry {\color{black}allowing for the full reconstruction of KDQs entering the weak values of $\hat{O}$}. It may prove relevant for the experimental revelation of strange correlators.

%%%%%%%%%%%%%%%%%%%%%%%%%
\section{Kirkwood-Dirac quasiprobabilities} 

Kirkwood-Dirac quasiprobabilities were introduced to describe the distribution of the outcomes from sequential measurements on a quantum system performed at consecutive times, without the concerned states of the system being disturbed by the  intermediate measurements. This objective has also been pursued by other measurement protocols developed in the last decade, such as those giving rise to non-demolition quasiprobabilities~\cite{Calarco1997,solinas2021,solinas2022,DeChiara2018,GherardiniTutorial}. Their peculiarity stems from the fact that the negativity of the corresponding distribution is a necessary and sufficient condition for the violation of macrorealism~\cite{solinas2024,SolinasReview2026}. KDQs, however, offer the advantage of directly providing a first-principles formulation of two-time quantum correlators, which also accounts for the noncommutativity of the involved operators.

To introduce the two-time KDQ framework, let us consider a quantum system initialized in the density operator $\hat{\rho}$, which is then consecutively measured at the consequent times $t_1$ and $t_2$ through the observables $\hat{O}_1(t_1)$ and $\hat{O}_2(t_2)$. In general, the operators $\hat{\rho}$, $\hat{O}_1(t_1)$, and $\hat{O}_2(t_2)$ do not commute with each other, i.e., they are incompatible~\cite{GherardiniTutorial}. Since KDQs describe the two-time statistics associated with measurement outcomes, the possible noncommutativity of $\hat{\rho}$, $\hat{O}_1(t_1)$, and $\hat{O}_2(t_2)$ implies that a KDQ does not obey to all the Kolmogorov axioms for classical probabilities. Nevertheless, KDQs sum to $1$, are linear in $\hat{\rho}$, and return the correct marginal distributions. In the context of two-time sequential quantum measurements, the term ``correct marginal distributions'' refers to the probability of measuring the observable $\hat{O}_1(t_1)$ or $\hat{O}_2(t_2)$ individually, without the system state being perturbed by any prior measurement. The nonclassical nature of KDQs manifests itself in the fact that they can take negative or even complex values. Each of these features can be interpreted as a loss of positivity of the corresponding statistics of the measurement outcomes.

In relation to a two-time statistics, a KDQ is defined by
\begin{equation}\label{eq:def_KDQ}
q_{s_1,s_2} = {\rm Tr}\left( \hat{\rho} \,  \hat{\Pi}_{s_1}(t_1) \,  \hat{\Pi}_{s_2}^{H}(t_2) \right),
\end{equation}
where $s_1$ and $s_2$ are the outcomes obtained from evaluating $\hat{O}_1$ and $\hat{O}_2$ at times $t_1$ and $t_2$, respectively~\cite{GherardiniTutorial}. In Eq.~\eqref{eq:def_KDQ}, the superscript $H$ denotes the evolution of a given observable in the Heisenberg representation, and $\hat{\Pi}_{s_i}$ ($i=1,2$) are the projectors of $\hat{O}_1(t_1)$ and $\hat{O}_2(t_2)$ that can be obtained from their spectral decomposition:
$\hat{O}_{i}(t_{i}) = \sum{s_{i}} o_{s_{i}}(t_{i}) \hat{\Pi}_{s_{i}}(t_{i})$. If we introduce $\Delta o_{s_1,s_2} = o_{s_2}(t_2) - o_{s_1}(t_1)$ as a generic difference between the measurement outcomes, then the probability distribution of $\Delta o$ in terms of the KDQ of Eq.~\eqref{eq:def_KDQ} is expressed as $P[\Delta o] = \sum_{s_1,s_2} q_{s_1,s_2} \delta(\Delta o - \Delta o_{s_1,s_2})$. The same information carried by the distribution $P[\Delta o]$ is contained in its complex Fourier transform, which is known as the characteristic function of $\Delta o$. Formally, it reads as~\cite{GherardiniTutorial}
\begin{equation}
\mathcal{G}(u) \equiv \int_{-\infty}^{+\infty} P[\Delta o] \, e^{i u \Delta o} \, d\Delta o = {\rm Tr}\left( \hat{\rho} \,  e^{-i u \hat{O}_1(t_1)} e^{i u \hat{O}^H_2(t_2)} \right),
\end{equation}
where $u$ is the (generally complex) variable  in terms of which the Fourier transform is performed. For system dynamics governed by a sudden quench transformation, the KDQ characteristic function $\mathcal{G}(u)$ takes the form of a generalized Loschmidt echo applied to the generic density operator $\hat{\rho}$~\cite{GherardiniTutorial,DonelliPRA2026}. This explains the capability of $\mathcal{G}(u)$ to be sensitive to state variations, due to changes in the Hamiltonian ruling the considered system or even to external perturbations.

We comment finally that a (Euclidean-)time evolution picture, parallel to that defining KDQ above, also suggests and justifies à posteriori the effectiveness of the strange correlators as a topology witness. Indeed, Eq.~\eqref{matper} can also be interpreted as correlator at the Euclidean time $\tau = 0$ between the evolved state $\ket{\Psi}$ from $\tau = - \infty$ to  $\tau = 0$ and the evolved state $\ket{\Omega}$ from $\tau = \infty$ to $\tau = 0$. If the two states have different topology and protected edges states characterizes $\ket{\psi}$, the theory at $\tau = 0$ is critical, then the correlators of $\hat{o}({\bf r})$ decay algebrically. This formal parallelism suggests a more deep relation between KDQ and strange correlators, investigated in more detail in the following sections.

%%%%%%%%%%%%%%%%%%%%%%%%%
\section{Strange correlators in terms of KDQs}

In this section, central for our work, we discuss how strange correlators can be expressed in terms of KDQ, {\color{black} interpreting it as a weak value of an operator that makes topological a phase, otherwise trivial.} While the scaling analysis of the strange correlator $s[\hat{o}]_{{\bf r},{\bf r^{\prime}}}$ is valid in principle for any topological system, we initially consider a textbook model of two-dimensional Chern (integer quantum Hall) insulator, displaying anomalous Hall response~\cite{bernevigbook,kotetesbook,notaqsh1}. This is the so-called {\it Bernevig-Hughes-Zhang} ({\bf BHZ}) model, a paradigmatic example of symmetry-protected topology. Starting from this discussion, other notable topological systems will be considered later on in the paper.    

\subsection{Two-dimensional Bernevig-Hughes-Zhang model}

The model is ruled by a two-dimensional single-particle Hamiltonian, diagonal in the quasi-momentum $\mathbf{k}$, which reads
\begin{equation}
\hat{H}^{\mathrm{(2D)}}_{\mathbf{k}} =
\begin{pmatrix}
\tilde{M}_{\mathbf{k}} & \sin(k_x) + i \sin(k_y) \\
\sin(k_x) - i\sin(k_y) & -\tilde{M}_{\mathbf{k}}
\end{pmatrix} \, ,
\label{BHZ_2D}
\end{equation}
where $\tilde{M}(\mathbf{k}) \equiv  M - 4 + 2 \, (\cos(k_x) + \cos(k_y))$, with $M$ denoting an effective mass parameter. The two-component spinor structure of Eq.~\eqref{BHZ_2D} can either originates from a sub-lattice or from orbital degrees of freedom, or it can be an effective representation of an inner spin degree of freedom for spin-orbit coupled systems. In the following, we will label the two spinor components with $A$ and $B$.

The Hamiltonian $\hat{H}^{\mathrm{(2D)}}_{\mathbf{k}}$ does not possess time-reversal invariance and, when considering a generic chemical potential, it belongs to the A-class of the (``tenfold-way'') classification of topological insulators and superconductors~\cite{altland1997,ludwig2009,kotetesbook}. Instead, at half-filling the same Hamiltonian is charge-conjugation invariant, i.e. $\hat{H}^{\mathrm{(2D)}}_{\mathbf{k}} = -\hat{\sigma}_1 \, \hat{H}^{\mathrm{(2D)} \, *}_{-\mathbf{k}} \, \hat{\sigma}_1$, where $\hat{\sigma}_1$ is the Pauli matrix along the $x$-axis. As a result, the energy bands display opposite energies at opposite momenta, and $\hat{H}^{\mathrm{(2D)}}_{\mathbf{k}}$ belongs to the D-symmetry class~\cite{altland1997,ludwig2009,kotetesbook}.

Closely related to the described symmetry content, the Hamiltonian $\hat{H}^{\mathrm{(2D)}}_{\mathbf{k}}$ has different phases depending on the value of $M$. For $M<0$ and $M>8$, the system is fully gapped and it corresponds to a standard band insulator with vanishing Chern number~\cite{bernevigbook}. Instead, for $0<M<4$ and $4<M<8$,  two topological gapped phases appear with Chern numbers $\pm 1$, thus displaying opposite Hall conductivities. These phases are separated by critical points at $M=0,4,8$, characterized by Dirac band-touching points. When open boundary conditions are adopted, the presence of a nontrivial topology results in the onset of chiral edge modes at the boundaries of the topological insulator phases~\cite{bernevigbook}.

\begin{figure}
\includegraphics[scale=0.22]{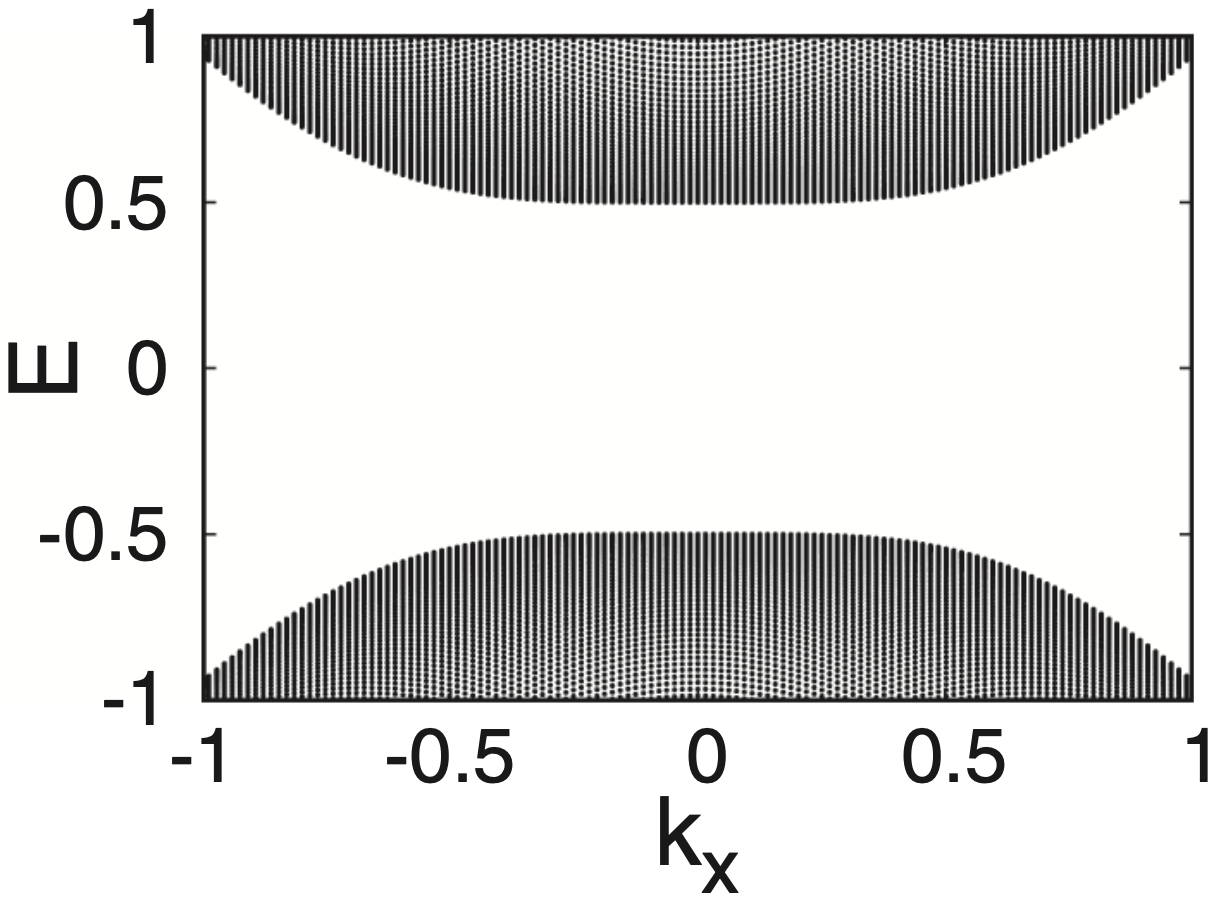}
\includegraphics[scale=0.22]{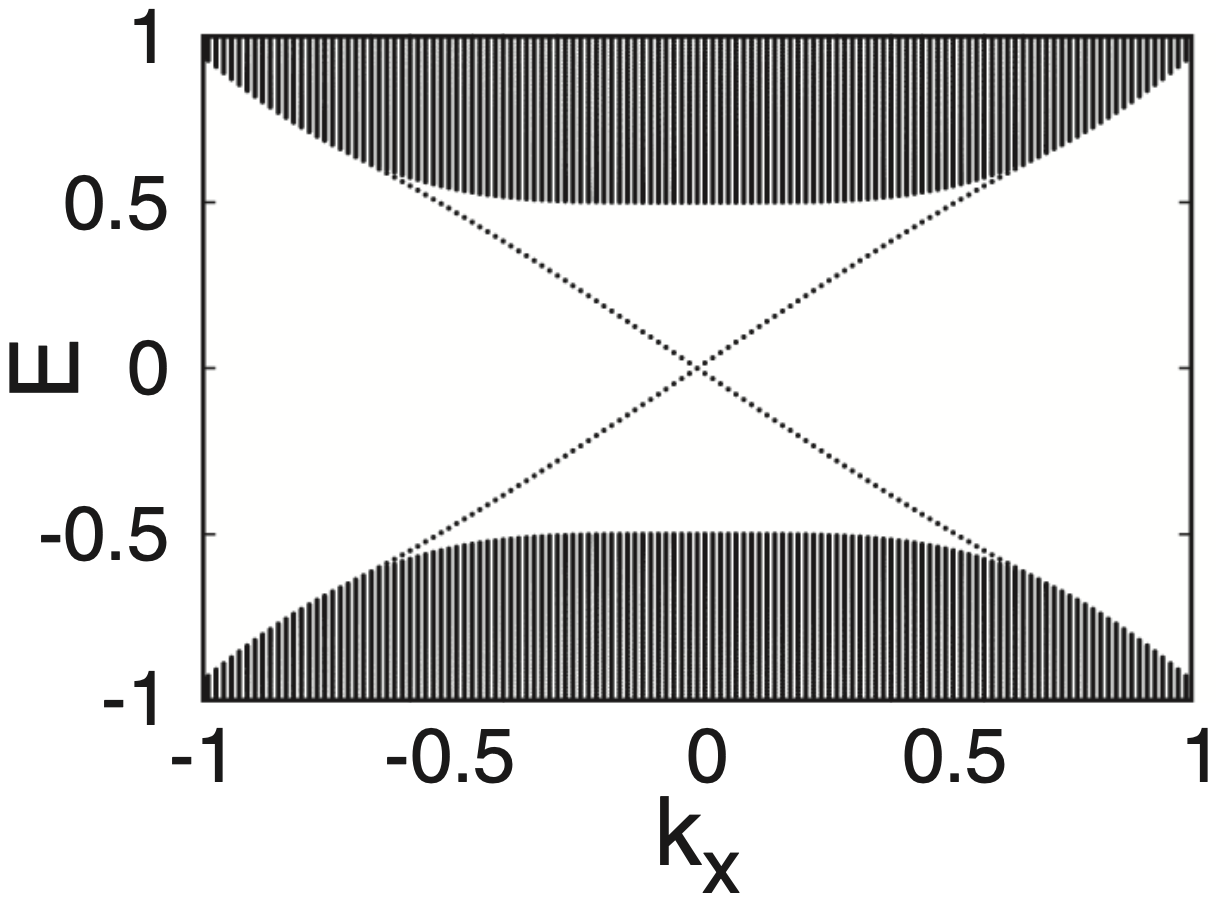}
\caption{Energy spectra of the Hamiltonian in Eq.~\eqref{BHZ_2D} at $M = 1$, for periodic boundary conditions (left panel) and with open boundary conditions along $y$ (right panel), respectively. In the right panel, gapless edge modes, topologically protected and interpolating between the two bands, are present.}
\label{plotspectra}
\end{figure}

The spectrum of $\hat{H}^{\mathrm{(2D)}}_{\mathbf{k}}$ at $M = 1$ is reported in Fig.~\ref{plotspectra}, for periodic boundary conditions (left panel) and with open boundary conditions along $y$ (right panel), respectively. In the first case, the spectrum displays two energy bands separated by an energy gap, with minimum at $\bf{k} = 0$; see \cite{lepori2022} and references therein. In the second case, the mentioned gapless edge modes, topologically protected and interpolating between the two bulk bands, are present.
{\color{black}The two limiting situations can be connected adiabatically, continuously linking an edge with its opposite one by an hopping with varying amplitude, from zero to the bulk value~\cite{lepori2022}. A detailed and general construction of two-dimensional topological insulators, making one-dimensional Luttinger chains interact, is reported in the supplementary material of \cite{xu2014}. From this construction, the presence of edge modes emerge naturally.}

The many-body ground-states of $\hat{H}^{\mathrm{(2D)}}_{\mathbf{k}}$ at half-filling are given by $|\Phi\rangle = \bigotimes_{i_{\mathbf{k}}} \hat{\eta}^{(-)\,\dagger}_{{\bf k}} |0\rangle_{i_{\mathbf{k}}}$, where $\Phi \in \{ \Omega, \Psi \}$, $|0\rangle_{i_{\mathbf{k}}}$ is the fermionic vacuum associated to the $i$-th momentum, and ${\bf k}$ spans the full BHZ Chern insulator. In the definition of $|\Phi\rangle$ and in all  the following, we label with $\hat{\eta}^{(\pm)}_{\bf k}$ ($\hat{\eta}^{(\pm)\,\dagger}_{\bf k}$) the annihilation (creation) operators of bulk quasiparticles at the top of the lowest- and at the bottom of the highest-energy band, assuming closed boundary conditions. In particular, 
\begin{equation}
\hat{\eta}^{(\pm)\,\dagger}_{{\bf k}} = a^{(\pm)}_{A,{\bf k}} \, \hat{c}^{\dagger}_{A, {\bf k}} + a^{(\pm)}_{B,{\bf k}} \, \hat{c}^{\dagger}_{B, {\bf k}},
\label{trasf}
\end{equation}
where the coefficients $a_{f,{\bf k}}^{(\pm)}$ are the elements of the eigenvectors of $\hat{H}^{\mathrm{(2D)}}_{\mathbf{k}}$, obtained from its diagonalization. Notice that the operators $\hat{\eta}^{(\pm)}_{{\bf k}}$ depend on $M$; however, throughout the paper, such a dependence on $M$ will be omitted for sake of brevity.

The Hamiltonian model in Eq.~\eqref{BHZ_2D} can be extended by adding an internal degree of freedom that doubles the dimension of the Hamiltonian matrix. The further  extension to three dimensions can be obtained via a suitable coupling of the system along the $z$ direction~\cite{bernevigbook,lepori2022}. Both in the two- and in the three-dimensional cases, the resulting time-reversal invariant Hamiltonian operators belong to the AII class of the tenfold-way classification of the topological insulators and superconductors~\cite{altland1997,ludwig2009,kotetesbook}, i.e., the same class of the quantum-spin Hall states. In the three-dimensional case, the main difference in the strange correlators are the scaling exponents, that change depending on the different scaling dimensions of the chosen bulk operators~\cite{lepori2022}.

\subsection{Decomposition of the strange correlators in terms of {\color{black}KDQs and interpretation as weak values}}\label{sec:decomposition}

We now consider the strange correlators between many-body ground-states of the Hamiltonian in Eq.~\eqref{BHZ_2D}, obtained for different values of the parameter $M$, i.e., $M_\Omega$ and $M_\Psi$~\cite{enlarge}. This choice is suggested by the requirement $\langle\Omega|\Psi\rangle \neq 0$,  at any finite size. Moreover, we assume periodic boundary conditions. 
Although the latter conditions are adopted for simplicity in the presentation, all the obtained results can be expressed in coordinate space without any complication or loss of generality. Finite systems with periodic conditions can be realized experimentally as well, say in ultracold atoms setups, via suitable electromagnetically-driven hyperfine transitions~\cite{boada2012,mancini2015}.

Here, the correct choice for the bilinear form of the operators $\hat{o}({\bf r})$ appearing in Eq.~\eqref{matper} is expressed in momentum space~\cite{lepori2022}:
\begin{equation}\label{eq:bilinear_momenta}
\hat{o}_{\bf k}\hat{o}_{\bf k^{\prime}} = \hat{\eta}^{(+)\,\dagger}_{\bf k^{\prime}} \hat{\eta}^{(-)}_{\bf k} \, \delta_{{\bf k},{\bf k}^{\prime}},
\end{equation}
where the operators $\hat{\eta}_{\bf k}$ refer to $\hat{H}^{\mathrm{(2D)}}_{\mathbf{k}}$ with $M = M_{\Psi}$ and ${\bf k} \approx 0$. The states with momenta around zero can become edge states as the boundary conditions are relaxed towards open conditions~\cite{lepori2022}, while the other states with greater momenta remain delocalized in the bulk.

Starting from Eq.~\eqref{eq:bilinear_momenta}, in order to analyze how the corresponding strange correlators relate to quasiprobabilities, we introduce the Hermitian part of $\hat{o}_{\bf k}\hat{o}_{\bf k^{\prime}}$, still in momentum space: 
\begin{equation}\label{defO}
\hat{O}_{\bf k} = \frac{1}{2}\Big( \hat{\eta}^{(+)\,\dagger}_{\bf k} \hat{\eta}^{(-)}_{\bf k} + \hat{\eta}^{(-)\,\dagger}_{\bf k} \hat{\eta}^{(+)}_{\bf k} \Big).
\end{equation}
Notice that, instead of Eq.~\eqref{defO}, one could equivalently consider even the operator $i \, ( \hat{\eta}^{(+)\,\dagger}_{\bf k} \hat{\eta}^{(-)}_{\bf k} - \hat{\eta}^{(-)\,\dagger}_{\bf k} \hat{\eta}^{(+)}_{\bf k} )/2$. From Eq. \eqref{defO}, we define the state
\begin{equation}
|\Upsilon_{\bf k}\rangle =
\hat{\eta}^{(+)\,\dagger}_{\bf k} \hat{\eta}^{(-)}_{\bf k} |\Psi\rangle 
\end{equation}
that represents the dynamical effect of bringing a spin excitation from the lowest energy-band, entirely filled and characterizing the state $|\Psi\rangle$, to the highest energy-band. Clearly, it holds that $\langle\Psi| \Upsilon_{\bf k} \rangle = 0$.
Substituting $\hat{O}_{\bf k}$ in the place of $\hat{o}_{\bf k}\hat{o}_{\bf k^{\prime}}$ within the strange correlator of Eq.~\eqref{matper} in momentum space, we can get the Hermitian part $s_{H}[\hat{o}_{\bf k}]$ of the strange correlator. Interestingly, the latter quantity is exactly equal to $s[\hat{o}_{\bf k}]$:
\begin{equation}
s_{H}[\hat{o}_{\bf k}] = s[\hat{o}_{\bf k}] =\frac{\langle\Omega|\Upsilon_{\bf k}\rangle}{\langle\Omega|\Psi\rangle}.
\label{defstrangemom}
\end{equation}

From Ref.~\cite{lepori2022}, it turns out that $s[\hat{o}_{\bf k}]$ has a different behavior around ${\bf k} = 0$, depending whether $|\Omega\rangle$ and $|\Psi\rangle$ share the same topology or not, namely if they are labeled or not by the same integer value of the proper topological index. From the one hand, if $|\Omega\rangle$ is topologically trivial but $|\Psi\rangle$ is not trivial, then for ${\bf k} \rightarrow 0$ it holds that
\begin{equation}\label{op1}
| s[\hat{o}_{\bf k}] |  \sim \frac{1}{|{\bf k}|^{\alpha}} \quad \text{with} \quad \alpha \geq 0 \,,
\end{equation}
where $|{\bf k}|$ is the modulus of ${\bf k}$ and $\alpha = 2 \, \eta_E$, where $\eta_E$ is the scaling dimension of the edge operators~\cite{lepori2022}. For the BHZ two-dimensional model given by Eq.~\eqref{BHZ_2D}, $|{\bf k}| = \sqrt{k_x^2 + k_y^2}$ and $\eta_E = \frac{D-1}{2} = \frac{1}{2}$. More in detail, both the terms of the ratio in Eq.~\eqref{defstrangemom} tend to zero as ${\bf k} \to 0$ 
(in parallel the system size $L$ diverges), and the numerator $\langle\Omega|\Upsilon_{\bf k}\rangle$ vanishes faster (see~\cite{lepori2022} and the supplemental material of \cite{xu2014}).

On the other hand, if $|\Omega\rangle$ and $|\Psi\rangle$ share the same topology (i.e., both states have trivial or non-trivial topology), then around ${\bf k} = 0$ we determine that
\begin{equation}\label{op2}
\left| s[\hat{o}_{\bf k}] \right| 
\sim |{\bf k}|^{\beta} \quad \text{with} \quad \beta \geq 0 \,.
\end{equation}
The exponent $\beta$ does not take a universal expression.

Being an Hermitian operator, $\hat{O}_{\bf k}$ can be expressed in terms of projectors: $\hat{O}_{\bf k} = \sum_{j=1}^{2}\lambda_j\hat{\Lambda}_{{\bf k},j}$. Here, $\lambda_1=1$, $\lambda_2=-1$ and the projectors $\hat{\Lambda}_{{\bf k},j}$ are equal to 
\begin{equation}
\hat{\Lambda}_{{\bf k},1} = |S^{(+)}_{\bf k}\rangle\!\langle S^{(+)}_{\bf k}| \quad \mathrm{and} \quad
\hat{\Lambda}_{{\bf k},2} = |S^{(-)}_{\bf k}\rangle\!\langle S^{(-)}_{\bf k}| \, ,
\end{equation}
where $|S^{(\pm)}_{\bf k}\rangle = \frac{1}{\sqrt{2}} \left( |\Psi\rangle \pm |\Upsilon_{\bf k}\rangle \right)$. As a result, 
\begin{equation}
\hat{O}_{\bf k} = 
|S^{(+)}_{\bf k}\rangle\!\langle S^{(+)}_{\bf k}| - |S^{(-)}_{\bf k}\rangle\!\langle S^{(-)}_{\bf k}| = |\Psi \rangle\!\langle \Upsilon_{\bf k} | + | \Upsilon_{\bf k} \rangle\!\langle \Psi| \,,
\end{equation} 
so that $\hat{O}_{\bf k}^2 = |\Psi \rangle\!\langle \Psi| + |\Upsilon_{\bf k} \rangle\!\langle \Upsilon_{\bf k} |$, $\hat{O}_{\bf k}^3 =\hat{O}_{\bf k}$ and
\begin{equation}\label{eq:exp_O_k}
e^{i u \hat{O}_{\bf k}} \ket{\Psi} = \cos(u) \, \ket{\Psi} + i \sin(u) \, \ket{ \Upsilon_{\bf k}}.
\end{equation}

We now introduce the density operators given by the outer product of $|\Psi\rangle$ and $|\Omega\rangle$ respectively:
\beq
\hat{\rho} = |\Psi\rangle\!\langle\Psi| \quad \mathrm{and} \quad
\hat{\Xi} = |\Omega\rangle\!\langle\Omega| \, ,
\eeq
 which are projectors and, in general, do not commute each other. The quantum correlator of $\hat{\Xi}$ and of the $j$-th projector $\hat{\Lambda}_{{\bf k},j}$  takes the expression \big($|S_{{\bf k},1}\rangle = |S^{(+)}_{\bf k}\rangle$ and $|S_{{\bf k},2}\rangle=|S^{(-)}_{\bf k}\rangle $\big):
\begin{eqnarray}
\label{eq:def_quasiprob}
q_{{\bf k},j} &=& {\rm Tr}\left( \hat{\rho} \, \hat{\Xi} \, \hat{\Lambda}_{{\bf k},j} \right) = \langle\Psi|\Omega\rangle \langle\Omega|\hat{\Lambda}_{{\bf k},j}|\Psi\rangle = 
\langle\Psi|\Omega\rangle \langle\Omega|S_{{\bf k},j}\rangle \langle S_{{\bf k},j}|\Psi\rangle = 
\nonumber \\
&=&  \frac{1}{2} \Big(|\langle\Psi|\Omega\rangle|^2 + (-1)^{(j+1)} \, \langle \Omega|\Upsilon_{\bf k}\rangle \langle\Psi|\Omega\rangle \Big) = \nonumber \\ 
&=& \frac{1}{2} \langle\Psi|\Omega\rangle  \Big(\langle\Omega|\Psi\rangle + (-1)^{(j+1)} \, \langle \Omega|\Upsilon_{\bf k}\rangle  \Big).
\end{eqnarray}
Here it is worth noting that, according to Eq.~\eqref{eq:def_KDQ}, ${\rm Tr}( \hat{\rho} \, \hat{\Xi} \, \hat{\Lambda}_{{\bf k},j} )$ in Eq.~\eqref{eq:def_quasiprob} represent two-time quantum correlators, built through a procedure of sequential non-projective measurements. In such a setting, $\hat{\Xi}$ and $\hat{\Lambda}_{{\bf k},j}$ have to be evaluated at two consecutive times separated by a sudden quench transformation. As explained above, in fact, $\hat{\Xi}$ and $\hat{\Lambda}_{{\bf k},j}$ are defined over two non-commuting bases, which map to each other through the operator $\hat{O}_{\bf k}$.

The correlators $q_{{\bf k},j}$ are KDQ~\cite{LostaglioQuantum2023,GherardiniTutorial,ArvidssonShukur2024review}, which can take both real negative values and imaginary ones, depending on the incompatibility of the operators $\hat{\rho}$, $\hat{\Xi}$ and $\hat{\Lambda}_{{\bf k},j}$. The second line of Eq.~\eqref{eq:def_quasiprob} makes explicit that the same parts can assume also negative values. Moreover, as $\sum_{j}\hat{\Lambda}_{{\bf k},j}=\hat{\mathbb{I}}$, with $\hat{\mathbb{I}}$ denoting the identity operator, it holds that $\sum_{j}q_{{\bf k},j}={\rm Tr}( \hat{\Xi} \, \hat{\rho} )$, that is the expectation value of $\hat{\Xi}=|\Omega\rangle\!\langle\Omega|$ with respect to $\hat{\rho}$, or vice-versa.

The strange correlator $s[\hat{o}_{\bf k}]$ can be recast directly as a sum of $\{q_{{\bf k},j}\}$. To do this, we define the overlap 
\begin{equation}
p_{\Xi} \equiv |\langle\Omega|\Psi\rangle|^2
\end{equation}
to measure $\hat{\Xi}$ in the correspondence of the quantum state $|\Psi\rangle$, such that
\begin{equation}\label{sumquasip}
s_{H}[\hat{o}_{\bf k}] = s[\hat{o}_{\bf k}] = \frac{1}{p_{\Xi}}\sum_{j}\lambda_j \, q_{{\bf k},j}\,.
\end{equation}
{\color{black}Eq.~\eqref{sumquasip}, together with Eq.~\eqref{defstrangemom}, exemplifies that the strange correlators $s[\hat{o}_{\bf k}]$ can be interpreted as weak values~\cite{LostaglioQuantum2023} of the operator $\hat{O}_{\bf k}$ [Eq.~\eqref{defO}] in momentum space. The decomposition of $s[\hat{o}_{\bf k}]$ in terms of KDQs is the core result of the paper.}

Being KDQs, $q_{{\bf k},j}$ are complex numbers, whose real and imaginary parts formally read as ${\rm Re}\,q_{{\bf k},j} = \frac{1}{2}{\rm Tr}\big( \hat{\rho} \, [ \hat{\Xi}, \hat{\Lambda}_{{\bf k},j} ]_+\big)$ and
${\rm Im}\,q_{{\bf k},j} = -\frac{i}{2}{\rm Tr}\big( \hat{\rho} \, [ \hat{\Xi}, \hat{\Lambda}_{{\bf k},j} ]_-\big)$, where $i$ is the imaginary unit, and $[\cdot,\cdot]_{\pm}$ denote the commutator and anti-commutator, respectively. More in detail,
\begin{eqnarray}
{\rm Re}\,s[\hat{o}_{\bf k}] = \frac{1}{p_{\Xi}}\sum_{j}\lambda_j \, {\rm Re}\,q_{{\bf k},j} &=& 
\frac{1}{2}{\rm Tr}\left( \hat{\rho} \left[ \hat{\Xi}, \hat{O}^{\prime}_{\bf k}  \right]_+ \right),\label{eq:real_part_sH} \\
{\rm Im}\,s[\hat{o}_{\bf k}] = \frac{1}{p_{\Xi}}\sum_{j}\lambda_j \, {\rm Im}\,q_{{\bf k},j} &=& 
-\frac{i}{2}{\rm Tr}\left( \hat{\rho} \left[ \hat{\Xi}, \hat{O}^{\prime}_{\bf k} \right]_- \right),\label{eq:imag_part_sH} 
\end{eqnarray}
where we have redefined $\hat{O}^{\prime}_{\bf k} = \hat{O}_{\bf k} / p_{\Xi}$. Since $\hat{O}_{\bf k} = |\Psi\rangle\!\langle\Upsilon_{\bf k}| + |\Upsilon_{\bf k}\rangle\!\langle\Psi|$, one has that
\begin{eqnarray}
[ \hat{\Xi},\hat{O}^{\prime}_{\bf k} ]_{\pm} &=& \frac{1}{\langle\Psi|\Omega\rangle} \Big( s[\hat{o}_{\bf k}]  |\Omega\rangle\!\langle\Psi| + |\Omega\rangle\!\langle\Upsilon_{\bf k}| \Big) \pm \mathrm{H.\,c.} \,.
\label{anticomm-extended}
\end{eqnarray}
Moreover, it is worth noting that the expression of ${\rm Im}\,s[\hat{o}_{\bf k}]$ given by Eq. \eqref{eq:imag_part_sH} can be further simplified in the small-time limit through linear response theory, as shown in the Appendix~\ref{sec:linear_response}.

In view of Eqs.~\eqref{eq:def_quasiprob} and \eqref{sumquasip}, the behavior of $s[\hat{o}_{\bf k}]$ around ${\bf k} = 0$ constrains that of KDQs $q_{{\bf k},j}$ in the same limit. Indeed, from Eqs.~\eqref{defstrangemom}-\eqref{op2}, it turns out that the leading behavior of KDQs around ${\bf k} = 0$ is
\begin{equation}\label{eq:scaling_KDQ_trivial}
q_{{\bf k},j} \sim \frac{1}{2} \, |\langle\Psi|\Omega\rangle|^2 
\end{equation} 
if $\ket{\Psi}$ is topologically trivial \big($\ket{\Omega}$ still assumed trivial\big), while
\begin{equation}\label{eq:scaling_KDQ_top}
q_{{\bf k},j} \sim  \frac{(-1)^{(j+1)}}{2} \, \langle \Omega|\Upsilon_{\bf k}\rangle \langle\Psi|\Omega\rangle 
\end{equation}
if $\ket{\Psi}$ is topologically non-trivial. 
{\color{black}The behaviors of Eqs.~(\ref{eq:scaling_KDQ_trivial})-(\ref{eq:scaling_KDQ_top}) is a direct consequence of the scaling of the ratio (\ref{defstrangemom}). Eqs.~\eqref{eq:scaling_KDQ_trivial}-\eqref{eq:scaling_KDQ_top} are also main results of this paper.} 

\begin{figure}
\includegraphics[scale=0.7]{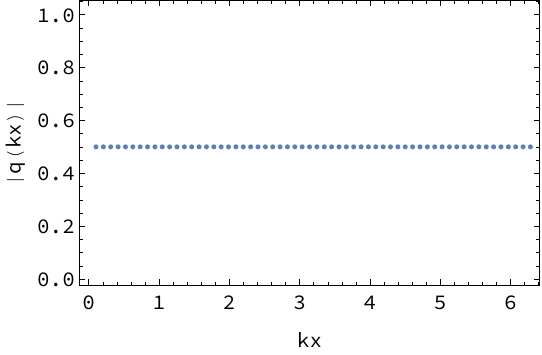}
\includegraphics[scale=0.7]{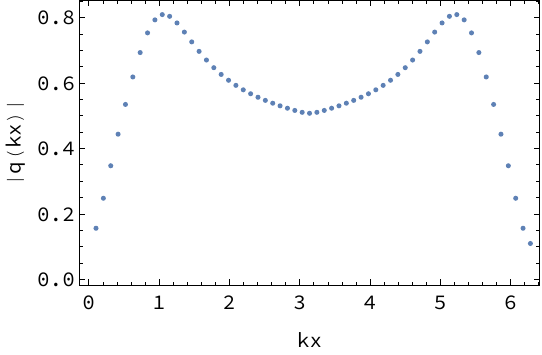}
\caption{{\color{black}Behavior of the quasiprobability modulus $|q_1\big(k_x , \frac{2 \pi}{L} \big)| \equiv |q(k_x)|$ ($L = 60$ being the linear dimension of the system) for $M_{\Omega} = -1$, $M_{\Psi} = - 1$ (left panel) and $M_{\Psi} = 1$ (right panel).}}
\label{plotqk}
\end{figure}

\begin{figure}
\includegraphics[scale=0.7]{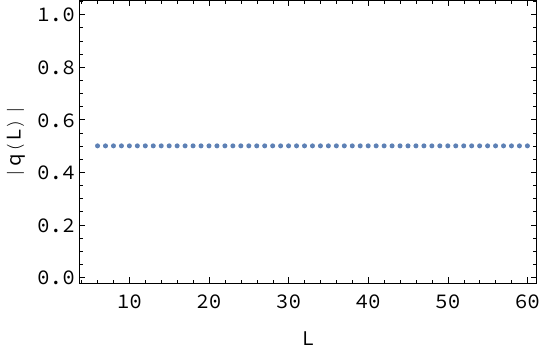}
\includegraphics[scale=0.7]{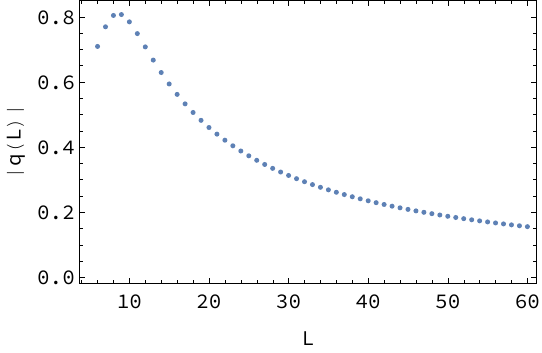}
\caption{{\color{black}Behavior of the quasiprobability modulus $|q_1 \left( \frac{2 \pi }{L}, \frac{2 \pi}{L} \right)| \equiv |q(L)|$, with $L$ denoting the linear dimension of the system, for $M_{\Omega} = -1$,  $M_{\Psi} = - 1$ (left panel) and $M_{\Psi} = 1$ (right panel).}}
\label{plotq}
\end{figure}

{\color{black}The described ${\bf k}$-dependence of the quasiprobability modulus $|q_1|$ is exemplified in Fig.~\ref{plotqk} (with fixed value of $L$). This also entails that, in coordinate space, the quasiprobabilities $q_{{\bf r},j}$---associated with Eqs.~(\ref{eq:scaling_KDQ_trivial})-(\ref{eq:scaling_KDQ_top}) by Fourier transform---are picked around ${\bf r} = 0$ when the state is non-topological, while they assume a non-trivial space-dependence if $\ket{\Psi}$ is topological. In Fig.~\ref{plotq}, instead, we report the $L$-dependence of the quasiprobability modulus $|q_1|$, with ${\bf k}$ set to $\frac{2 \pi }{L}$. Interestingly, in the topological regime, the largest value of $|q_1|$ is reached for a finite $L$, which depends on the considered momentum ${\bf k}$.}

The reconstruction of quasiprobabilities is in general enabled by a procedure based on quantum interferometry (discussed in detail below in Sec.~\ref{sec:interferometry}), in contrast to the reconstruction of the strange correlators $s[\hat{o}]_{{\bf r}, {\bf r^{\prime}}}$ and $s[\hat{o}_{\bf k}]$, for which no method is currently available.

\subsection{Further developments}

The described construction for the model in Eq.~\eqref{BHZ_2D} can be repeated for various systems, both with symmetry-protected and with genuine topological orders. For instance, one can also consider the \emph{Affleck-Kennedy-Lieb-Tasaki} ({\bf AKLT}) chain~\cite{aklt,lepori2022}, whose Hamiltonian is defined by
\begin{equation}
\hat{H}_{\mathrm{AKLT}} = \sum_{j} \hat{{\bf s}}_j \cdot \hat{{\bf s}}_{j+1} + \frac{1}{3} \left( \hat{{\bf s}}_j \cdot \hat{{\bf s}}_{j+1} \right)^{2}.
\end{equation}

The ground-state of $\hat{H}_{\mathrm{AKLT}}$ is a valence bond solid, with a single bond connecting every neighboring pair of sites. More in detail, each spin-1 $\hat{{\bf s}}_j$ can be decomposed in terms of two spins $\frac{1}{2}$ degrees of freedom, projected on a triplet state; in turn, a spin $\frac{1}{2}$ on a site is organized in a singlet together with a spin $\frac{1}{2}$ on the neighboring site.
In this way, if periodic boundary conditions are assumed, the AKLT chain has a unique ground-state, $\ket{\mathrm{GS}}$, fully characterized by the mentioned singlets. Instead, if open boundaries are assumed, the first and last spin 1 of the chain have only a single neighbor, leaving one of their constituent spins $\frac{1}{2}$ unpaired. As a result, the ends of the chain behave like free spins $\frac{1}{2}$. For finite chains, these edge states mix in a singlet and a higher-energy triplet states, similarly as for the closed chain, while, as the size increases, the edge states decouple exponentially, leading to a ground-state manifold that is four-fold degenerate. Therefore, the edge states are massless in the thermodynamic limit.
In this scheme, the ${\bf s}_j^{(\pm)}$ operators, applied at the end of a finite-size chain, create (+) or destroy (-) edge excitations with vanishing scaling dimension. More in detail, in a finite-size chain with periodic boundary conditions, a spin $\frac{1}{2}$ singlet is transformed to a general combination of the triplet states, i.e.,
\begin{equation}
\frac{ |\!\uparrow\downarrow\rangle - |\!\downarrow \uparrow\rangle }{\sqrt{2}} \longrightarrow \alpha |\!\uparrow \uparrow\rangle + \beta |\!\downarrow \downarrow\rangle + \gamma \frac{ |\!\uparrow \downarrow\rangle + |\!\downarrow \uparrow\rangle}{\sqrt{2}}, 
\end{equation}
with $|\alpha|^2 + |\beta|^2 + |\gamma|^2 = 1$, by means of the local (non-unique) operator
\begin{equation}
\hat{S} = \frac{\alpha}{\sqrt{2}} \left( \hat{\mathbb{I}} \otimes \hat{s}^{(+)} - \hat{s}^{(+)} \otimes \hat{\mathbb{I}} \right) + \frac{\beta}{\sqrt{2}} \left( \hat{s}^{(-)} \otimes \hat{\mathbb{I}} - \hat{\mathbb{I}} \otimes \hat{s}^{(-)} \right) + \frac{\gamma}{\sqrt{2}} \left( \hat{s}_z \otimes \hat{\mathbb{I}} - \hat{\mathbb{I}} \otimes \hat{s}_z \right) ,
\end{equation}
where $\hat{s}_z = \frac{\sigma_z}{2}$.
The operator $\hat{S}$ is applied on the adjacent sites that are linked to the bond that we relax towards open boundary conditions.
Therefore, $\hat{S}$ creates the lowest energy excitations above the unique ground-state $\ket{\mathrm{GS}}$. This construction explains the behavior 
\begin{equation}
\langle \Omega | \hat{{\bf s}}_i^{(+)} \hat{{\bf s}}_j^{(-)} | \Psi \rangle \to \mathrm{constant} 
\end{equation}
as $|i-j| \to \infty$, obtained for the closed chain in Refs.~\cite{xu2014,lepori2022}.

We can express straightforwardly the strange correlator for the AKLT chain in terms of KDQs by identifying $\ket{\Psi}$ with $\ket{\mathrm{GS}}$ ($\ket{\Psi} = \ket{\mathrm{GS}}$), such that
\begin{equation}
\ket{\Upsilon_{\bf k}} = \left( \hat{S} +\hat{S}^{\dagger} \right) \ket{\mathrm{GS}}.
\end{equation}

We also mention the further example of the {\it Laughlin states}, predominant in the context of the fractional quantum Hall effect.  For these states, the role of $\hat{\eta}^{(+)}_{\bf k}$ and $\hat{S}$ is played by the vertex operators that create elementary $e/3$ fractional excitations above the ground state~\cite{teokane,fradkinbook,lepori2022}.

We conclude this section discussing the cases where, instead of a pure state $\ket{\psi}$, a mixed state $\hat{\rho}$ is considered to probe the topology. A typical case is the \emph{thermal mixed state}. The formalism used in the present work makes no distinction whether the probe state is a pure or mixed state. Moreover, the capability of detecting topology should not change either. In fact, in Ref.~\cite{mera2017}, it has been shown that no phase transition between different topologies is driven by a finite temperature, unless the parameters of the considered Hamiltonian host an explicit dependence on the temperature itself, such as to induce a transition, typically on the ground-state. Therefore, taking into account the ground-state (assumed unique) of the many-body system as the probe state reveals sufficient à priori.\\ 
In a real experiment, however, one has to face with a non-zero temperature and instead of having $\hat{\rho} = |\Psi\rangle\!\langle\Psi|$, a thermal density matrix has to be considered: 
\begin{equation}
    \hat{\rho} = \sum_{i}f_{|i\rangle,T} |i\rangle\!\langle i| = f_{|\Psi\rangle,T} |\Psi\rangle\!\langle\Psi| + \sum_{e}f_{|e\rangle, T} |e\rangle\!\langle e|.
    \label{eq:mixed_state}
\end{equation}
In Eq.~\eqref{eq:mixed_state}, $|e\rangle$ denotes the excited energy eigenstates of the system Hamiltonian above the ground-state $|\Psi\rangle$; instead, $f_{|i\rangle, T}$ are the Gibbs weights (properly normalized to 1) of the thermal density matrix in the presence of the temperature $T$. In such a context, Eq.~\eqref{eq:def_quasiprob} becomes:
\begin{equation}
\label{eq:def_quasiprob_exc}
q_{{\bf k},j} = {\rm Tr}\left( \hat{\rho} \, \hat{\Xi} \, \hat{\Lambda}_{{\bf k},j} \right) = \frac{1}{2} \Big( f_{|\Psi\rangle,T} \langle\Psi|\Omega\rangle + \sum_{e} f_{|e\rangle,T} \langle e|\Omega\rangle \langle \Upsilon_{\bf k} | e \rangle \Big) \Big(\langle\Omega|\Psi\rangle + (-1)^{(j+1)} \, \langle\Omega|\Upsilon_{\bf k}\rangle \Big).
\end{equation}
Notice that for the BHZ model, $|\langle \Upsilon_{\bf k} |e^*\rangle| \to 1$ if $|e^*\rangle$ is the lowest exited state.
In conclusion, a nontrivial momentum dependence still develops primarily from the second factor of the right-hand side of Eq.~\eqref{eq:def_quasiprob_exc} due to the ground-state contribution; a similar argument holds for Eq.~\eqref{eq:scaling_KDQ_top}.

\begin{comment}
Since the excited eigenstates are topologically trivial, as well as $|\Upsilon_{\bf k} \rangle$ (that can be written in their terms), the argument of the sum is smooth as ${\bf k} \to 0$. 
\end{comment}
 
%%%%%%%%%%%%%%%%%%%%%%%%%
\section{Strategy towards experimental revelation}
\label{sec:interferometry}

In this section, we describe how expressing the strange correlators in terms of KDQs suggests a revelation strategy based on quantum interferometry that may find experimental applications. Perhaps, this aspect is not the main novelty of our paper, as instead is the emerging link between strange correlators and KDQs within the framework of sequential (non-projective) measurements. In fact, more direct strategies to reveal topologically-protected edge states are available so far in the current literature, see e.g.~\cite{vonk1980,vklitzinglec,foxon1988,hasan2009,chen2009,yazdani2012,konig2007,mancini2015}. Nonetheless, given that interferometry of topological quantum phases for detecting topological orders is an emerging tool~\cite{GrusdtNatComm2016}, we think that addressing explicitly the steps of our revelation strategy is highly relevant for future implementations.

As proposed in Refs.~\cite{LostaglioQuantum2023,GherardiniTutorial} and experimentally realized in \cite{hernandez2024Interfero}, the real and imaginary parts of the characteristic function of a KDQ distribution, under a time-dependent unitary dynamics, can be measured directly using an interferometric scheme.

In our case-study with topological quantum states, given the quasiprobability $q_{{\bf k},j} = {\rm Tr}( \hat{\rho} \, \hat{\Xi} \, \hat{\Lambda}_{{\bf k},j} )$---special kind of a two-point quantum correlator as explained in Sec.~\ref{sec:decomposition}---we introduce the 2-index quasiprobability
\begin{equation}
    q_{{\bf k},(i,j)} \equiv {\rm Tr}\left( \hat{\rho} \, \hat{\Xi}_i \, \hat{\Lambda}_{{\bf k},j} \right),
\end{equation}
where $\hat{\Xi}_1 = \hat{\Xi} = |\Omega \rangle\!\langle \Omega|$ and $\hat{\Xi}_2 = \hat{\Xi}^{\perp} = \mathbb{I} - |\Omega \rangle\!\langle \Omega|$. With this definition, $\sum_{i,j} q_{{\bf k},(i,j)} = 1$ and $q_{{\bf k},j}=q_{{\bf k},(1,j)}$. We also consider the Hermitian observable 
\begin{equation}
\hat{O}_{\Xi} \equiv \sum_{i=1}^{2}\lambda_{\Xi_i} \, \hat{\Xi}_i = 2  \, |\Omega \rangle\!\langle \Omega | - \mathbb{I},
\end{equation}
with $\lambda_{\Xi_i}$ real numbers ($\lambda_{\Xi_1}=1$ and $\lambda_{\Xi_2}=-1$). 
The quasiprobabilities $q_{{\bf k},(i,j)}$ define the distribution
\begin{equation}
    {\rm Prob}(\Delta o) = \sum_{i,j}q_{{\bf k},(i,j)}\,\delta\big( \Delta o - (\lambda_j - \lambda_{\Xi_i}) \big)
\end{equation}
that describes the statistics of the measurement outcomes of the differences $\Delta o_{i,j} \equiv \lambda_j - \lambda_{\Xi_i}$, where we recall that $\lambda_j$ are the eigenvalues that spectrally decompose $\hat{O}_{\bf k}$. Associated to the quasiprobability distribution ${\rm Prob}(\Delta o)$, it is worth considering the characteristic function
\begin{equation}
    \mathcal{G}_{{\bf k}}(u) = {\rm Tr}\left( \hat{\rho} \, e^{-i u \hat{O}_{\Xi}} e^{i u \hat{O}_{\bf k}} \right).
\end{equation}

A strategy to achieve the reconstruction of the full quasiprobability distribution ${\rm Prob}(\Delta o)$ is through the direct measurement of the characteristic function $\mathcal{G}_{{\bf k}}(u)$, using quantum interferometry~\cite{LostaglioQuantum2023,GherardiniTutorial,hernandez2024Interfero}. In fact, collecting measurements of the real and imaginary parts of $\mathcal{G}_{{\bf k}}(u)$, with $u$ taking the dimension of a time, ${\rm Prob}(\Delta o)$ is reconstructed by performing numerically the inverse Fourier transform of the measured characteristic function. Then, from ${\rm Prob}(\Delta o)$, also the quasiprobabilities $q_{{\bf k},(i,j)}$ and $ q_{{\bf k},j}$ are obtained.

\begin{figure}[t]
\includegraphics[scale=1.3]{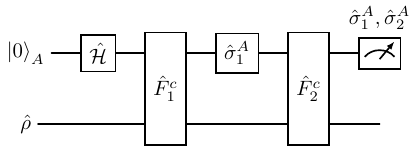}
\caption{
Pictorial representation of the interferometric scheme that realizes the measurement of the real and imaginary parts of $\mathcal{G}_{{\bf k}}(u)$.
}
\label{fig:interferometer}
\end{figure}

As depicted in Fig.~\ref{fig:interferometer}, the interferometric scheme makes use of an auxiliary system $A$ that, without loss of generality, can be taken as a qubit. This system is initialized in the ground state $|0\rangle_A$ to which the Hadamard gate $\hat{\mathcal{H}}$ is applied on, while the system under scrutiny is represented by  the density matrix  $\hat{\rho}$. The interferometric scheme is based on the implementation of two conditional quantum gates~\cite{GherardiniTutorial}, $\hat{F}_{1}^{c}$ and $\hat{F}_{2}^{c}$, that, for our specific case-study, are
\begin{eqnarray}
    \hat{F}_{1}^{c} &=& e^{-iu\hat{O}_{\Xi}} \otimes |0\rangle_{A}\langle 0| + \hat{\mathbb{I}} \otimes |1\rangle_{A}\langle 1|\label{fport1} \\ 
    \hat{F}_{2}^{c} &=& \hat{\mathbb{I}} \otimes |0\rangle_{A}\langle 0| + e^{iu\hat{O}_{\bf k}} \otimes |1\rangle_{A}\langle 1|,\label{fport2}
\end{eqnarray}
where $e^{iu\hat{O}_{\bf k}}$ is defined in Eq.~\eqref{eq:exp_O_k}, while $e^{iu\hat{O}_{\Xi}}$, when applied to $|\Psi\rangle$, returns: 
\begin{equation}
e^{ i u \hat{O}_{\Xi}} \ket{\Psi} = e^{- i u} \, \Big( \hat{\mathbb{I}} + (e^{ i 2 u} -1) | \Omega \rangle\!\langle \Omega | \Big) \ket{\Psi} 
= e^{- i u} \, \Big( \ket{\Psi} + (e^{i 2 u} - 1) \langle \Omega \ket{\Psi}\ket{\Omega} \Big) \, .
\label{trasf2}
\end{equation}
After the application of the second conditional gate $\hat{F}_{2}^{c}$, the state of the auxiliary system is measured with respect to the measurement observables $\hat{\sigma}_{1}^{A}$ and $\hat{\sigma}_{2}^{A}$, with $\hat{\sigma}_{1}^{A}, \hat{\sigma}_{2}^{A}$ denoting the Pauli matrices along the $x$- and $y$-axis, respectively. Indeed, as explained in Ref.~\cite{GherardiniTutorial}, the expectation values of $\hat{\sigma}_{1}^{A}, \hat{\sigma}_{2}^{A}$ at the end of the interferometric scheme are equal to the real and imaginary parts of $\mathcal{G}_{{\bf k}}(u)$ for the given time $u$: ${\rm Re}\,\mathcal{G}_{{\bf k}}(u) = \langle \hat{\sigma}_{1}^{A}(u)\rangle$ and ${\rm Im}\,\mathcal{G}_{{\bf k}}(u) = \langle \hat{\sigma}_{2}^{A}(u)\rangle$. In order to reconstruct the quasiprobability distribution ${\rm Prob}(\Delta o)$, the interferometric procedure described above must be repeated for many values of $u$ (taken as real numbers with the dimension of time): the greater the number of $u$, the more accurate is the reconstruction of ${\rm Prob}(\Delta o)$.

The general structure of the operations in Eqs.~(\ref{fport1})-(\ref{fport2}), controlled by the ancilla and therefore entangling with the many-body state $\ket{\psi}$, {\color{black}appears not trivial to be realized experimentally on extended quantum systems. However, it is known that some topologically not-trivial systems can be effectively encoded in single qubits or qudits. Here, we provide a realistic physical example, involving a two-level system hosting not-trivial topology, where similar controlled gates can be performed.} This example stems from the BHZ model in Eq.~\eqref{BHZ_2D}.

In the contest of the Floquet theory~\cite{vieban2020}, it is known that the topology exhibited by the BHZ model can be realized in time, formally replacing~\cite{halperin2017}
\begin{equation}\label{eq:floquet}
k_x \to \omega_x \, t + \phi_x \,,
\end{equation}
where $t$ labels the time, {\color{black} and $\omega_x$ is the angular frequency of time-dependent drivings with amplitude $\Gamma_x$ and phase $\phi_x$. Similar notations as in Eq.~\eqref{eq:floquet} hold for the other space directions.} In this way, $\hat{H}^{\mathrm{(2D)}}_{\mathbf{k}}$ in Eq.~\eqref{BHZ_2D} becomes
\begin{equation}
\hat{H}^{\mathrm{(2D)}}(t) = (M - 4 \, \Gamma_z) \, \hat{\sigma}_{z} + \Gamma_x  \sin\big(\omega_x  t + \phi_x \big) \, \hat{\sigma}_x + \Gamma_y  \sin\big(\omega_y  t + \phi_y \big) \,  \hat{\sigma}_y +  2 \, \Gamma_z   \big(\cos(\omega_x  t + \phi_x) + \cos(\omega_y  t + \phi_y) \big) \hat{\sigma}_{z} \, .
\label{BHZt}
\end{equation}
{\color{black} For simplicity, one can further set $\phi_y = 0$.}
The topology for the system in Eq.~\eqref{BHZt} is encoded in the structure of the related Floquet bands~\cite{demler2010}, and occurs for the same range of $M$ as for Eq.~\eqref{BHZ_2D}. A customary manifestation of this topology is the time counterpart of the Thouless pumping~\cite{thouless1983}, realized as a transfer of energy between the modes $\omega_x$ and $\omega_y$~\cite{halperin2017}.

The topology for the Hamiltonian of Eq.~\eqref{BHZt} can be revealed via the same interferometric protocol described above, {\color{black}which can be realized with currently-available techniques. In fact, the conditional quantum gates in Eqs.~(\ref{fport1})-(\ref{fport2}) can be performed by controlled operations entangling two two-level systems: the first one is the driven system of Eq.~\eqref{BHZt}, while the second one is an auxiliary (ancilla) system, controlling the operations and on which the real and imaginary part of $\mathcal{G}_{{\bf k}}$ are encoded for any $k$. Concerning the first system, $\ket{\Omega} = (0,1)$ can be chosen as the lowest energy eigenstate in the limit $M \to \infty$, while $\ket{\Psi}$ can be assumed at fixed $\phi_x$ as the lowest Floquet eigenstate (to be calculated along the standard approach described e.g.~in \cite{vieban2020,halperin2017}). In turn, this state coincides with the lowest instantaneous eigenstate at $t=0$ in the adiabatic-evolution limit~\cite{wauters2018}, a fact that indicates how to prepare $\ket{\Psi}$ in a possible experiment. With these choices, the overlap $\langle\Omega \ket{\Psi}$ (appearing in the transformation of Eq.~\eqref{trasf2}) is known, with the result that the transformation in Eq. (\ref{trasf2}) can be effectively performed. In Eq.~\eqref{trasf2}, $u$ has to be considered as a second time-dependent parameter, varying on a scale much larger than $\frac{\hbar}{\Gamma}$ and $\frac{2 \pi}{\omega_x}, \frac{2\pi}{\omega_y}$.

There are various ways to realize the model in Eq.~\eqref{BHZt}, as well as the operations of Eqs.~\eqref{fport1} and \eqref{fport2}.
A first possibility concerns neutral ultracold atoms, where one has to implement selective population of hyperfine levels and strong Rydberg blockade, required for fast controlled operations~\cite{zoller2000}.} Secondarily, 
as described in \cite{hernandez2024Interfero}, conditional gates similar to those of Eqs.~(\ref{fport1})-(\ref{fport2}) have been recently realized by entangling the electronic (the driven qubit under scrutiny) and nuclear (the auxiliary system) spins of a single nitrogen-vacancy (NV) center~\cite{cappellaro2016}. This has been achieved by exploiting the electron-nucleus interaction 
$\hat{H}_I = A \, \hat{\sigma}_z \otimes \hat{\sigma}_z$, where $A$ is the hyperfine constant that describes the coupling between the nuclear and the electronic spins. The hyperfine constant $A$ is assumed much larger than the amplitudes of the fields that drive the qubit system, a requirement similar to the strong Rydberg blockade {\color{black} in ultracold atoms. Finally, we mention that similar conditioned operations have been described recently in the context of molecular nanomagnets and between interacting qudit-manifolds (where two levels can be still selectively populated)~\cite{macedonio2026}. In such contexts, optimal control~\cite{rossignolo2022} has been exploited; it could be a relevant possibility also for our purposes.}

%%%%%%%%%%%%%%%%%%%%%%%%%
\section{Conclusions and perspectives}

In this work, we have discussed how strange correlators, theorized as an efficient witness for topological matter, can be expressed using Kirkwood-Dirac quasiprobabilites (KDQs). This fundamental link realizes in the decomposition of the strange correlators in terms of two-time quantum correlators, {\color{black}and allows for a direct interpretation of strange correlators as weak values. Such interpretation} involves introducing a sudden quench transformation that maps an initial trivial quantum state into a topologically non-trivial one. Our derivation clarifies the microscopic nature of the strange correlators, motivating further their ability to identify topological quantum features.

In recent years, many methods have been identified for the experimental measurement of quasiprobabilities~\cite{solinas2021,LostaglioQuantum2023,hernandezPRR2024,WagnerQST2024,GrusdtNatComm2016,hernandez2024Interfero,BizzarriQST2025}, including the Kirkwood-Dirac ones. Following this trend, and thanks to the direct link between strange correlators and KDQs, we have proposed a measurement scheme based on quantum interferometry. Although promising, the effective application of such a scheme to realistic topological systems requires further investigations. {\color{black}It would also be worth studying viable routes towards realizing weak measurements of the changing-topology operator in an actual experimental setting.

Finally, we mention the perspective to measure quasiprobabilities in extended quantum systems. For this aim, realizations on ultracold atoms platforms, as in Refs.~\cite{jotzu2014,mancini2015}, appear again particularly suitable. In similar settings, the effective role of the auxiliary system could be played by a two-state boson condensate with filling $\geq 1$. Suitable controlled operations could be achieved exploiting a strong repulsive interaction between a state of the auxiliary system and the fermions that comprise the topological system under scrutiny.}

%%%%%%%%%%%%%%%%%%%%%%%%%
\section*{ACKNOWLEDGEMENTS}

The authors are pleased to thank {\color{black} Giuseppe Bevilacqua,} Michele Burrello, and Simone Paganelli for fruitful discussions.
L.L. and S.G. acknowledge financial support by a project funded under the National Recovery and Resilience Plan (NRRP), Mission 4 Component 2 Investment 1.3 - Call for tender No. 341 of 15/03/2022 of Italian Ministry of University and Research funded by the European Union -- NextGenerationEU, award number PE0000023, Concession Decree No. 1564 of 11/10/2022 adopted by the Italian Ministry of University and Research, CUP D93C22000940001, Project title ``National Quantum Science and Technology Institute'' (NQSTI), spokes 2 and 3.

%%%%%%%%%%%%%%%%%%%%%%%%%%%%%%%%%%%%%%%%%%%

%%%%%%%%%%%%%%%%%
\newpage

\appendix

\section{Linear response theory applied to quantum topology probing}\label{sec:linear_response}

Kirkwood-Dirac quasiprobabilities have a direct link with the linear response theory~\cite{LostaglioQuantum2023}, which allows us to determine a simplified expression of the imaginary part of the strange correlator, as given by Eq.~\eqref{eq:imag_part_sH}, in the small-time limit.

In order to illustrate this fact, we start from the known Kubo formula~\cite{GV} in the time interval $[t_0,t]$:
\begin{equation}\label{Kubo}
\langle X(t) \rangle = \left\langle X(0) \right\rangle_0 - \frac{i}{\hbar}\int_{t_0}^{t} \left\langle \, \left[ X(t), V \left(t'\right)\right] \, \right\rangle_{0} \, dt',
\end{equation}
where $X(t)$ stands for a generic observable in the interaction picture, and $V(t)$ is a Hamiltonian perturbation. It is worth noting that in Eq.~\eqref{Kubo} the averages $\langle\cdot\rangle_0$ are made on the unperturbed state, while the standard average $\langle\cdot\rangle$ on the left-hand side is performed with respect to the current state of the system at time $t$. On the right-hand-side of Eq.~\eqref{Kubo} ,we set $X(0) = \hat{\Xi} = |\Omega\rangle\!\langle\Omega|$, such that
\begin{equation}\label{oprabi}
X(t) = e^{i H_0 (t-t_0) } \, |\Omega\rangle\!\langle\Omega| \, e^{-i H_0 (t-t_0) },
\end{equation}
where $H_0$ denotes the unperturbed Hamiltonian whose ground state is a topologically non-trivial pure state $\ket{\Psi}$, as in Eq.~\eqref{BHZ_2D} for $0 < M < 4$. Moreover, $\hat{\rho} = |\Psi\rangle\!\langle\Psi|$ and the Hamiltonian perturbation (in momentum space) is defined as  
\begin{equation}\label{pert}
V({\bf k}, t) = V_0 \, e^{i H_0 (t-t_0) } \, \hat{O}_{{\bf k}} \, e^{-i H_0 (t-t_0) } \, \theta(t-t_0) \, ,
\end{equation}
with ${\bf k} \to 0$. In Eq.~\eqref{pert}, $V_0$ is the energy amplitude of the perturbation and $\theta(\cdot)$ denotes the Heaviside step function. If $V_0$ is much lower that the norm of $H_0$, then the linear response regime is respected. 
In this regime, combining Eqs.~\eqref{eq:real_part_sH}-\eqref{eq:imag_part_sH}, \eqref{Kubo}, and \eqref{pert}, we find:
\begin{eqnarray}\label{eqlin}
\frac{|\langle \Psi(t)|\Omega\rangle|^2 - |\langle \Psi |\Omega \rangle|^2}{\langle\Omega|\Psi\rangle} &=& \frac{V_0}{\Delta} \left[ i \sin\left( \frac{\Delta}{\hbar} (t-t_0) \right) \langle\Psi| \, [\hat{\Xi}, \hat{O}^{\prime}_{\bf k}] \, |\Psi\rangle + \left( \cos\left( \frac{\Delta}{\hbar}(t-t_0) \right) -1 \right) \langle\Psi| \, \{ \hat{\Xi}, \hat{O}^{\prime}_{\bf k} \} \, |\Psi\rangle \right] = \nonumber \\
&=& 2 \, \frac{V_0}{\Delta} \left[ -\sin\left( \frac{\Delta}{\hbar} (t-t_0) \right) \, {\rm Im} \,s[\hat{o}_{\bf k \to 0}] + \left( \cos\left( \frac{\Delta}{\hbar}(t-t_0) \right) -1 \right) \, {\rm Re}\,s[\hat{o}_{\bf k \to 0}] \right],
\end{eqnarray}
where $\ket{\Psi(t)}$ denotes the evolution of $\ket{\Psi}$ after the perturbation in Eq.          \eqref{pert} is switched on. Moreover, $\Delta$ is the energy gap between $\ket{\Upsilon_{{\bf k} \to 0}}$ and $\ket{\Psi}$ in $H_0$; $\Delta$ is such that $H_0 \, \hat{O}_{{\bf k} \to 0} |\Psi\rangle =  H_0 \, |\Upsilon_{{\bf k} \to 0} \rangle = \Delta|\Upsilon_{{\bf k} \to 0} \rangle$, while $H_0 |\Psi\rangle \equiv 0$.

From Eq.~\eqref{eqlin}, we evince the presence of Rabi oscillations between $\ket{\Psi}$ and $\ket{\Upsilon_{{\bf k} \to 0}}$, with period $T = 2 \pi \frac{\hbar}{V_0}$. Such oscillations originate from $H_0$, entering both $X(t)$ and $V({\bf k}, t)$. In particular, in the two-level space composed by the states $\Psi$ and $\ket{\Upsilon_{{\bf k} \to 0}}$, the operator $X(t)$ in Eq.~\eqref{oprabi} reads as
\begin{equation}
X(t)  = 
\begin{bmatrix}
0  &  e^{i \Delta (t-t_0) }  \\
e^{- i \Delta (t-t_0) }  &  0
\end{bmatrix}, 
\end{equation}
while the perturbation in Eq.~\eqref{pert} makes a quasiparticle to evolve from the lower to the upper band, effectively inducing the oscillation $\ket{\Psi}\to \ket{\Upsilon_{{\bf k} \to 0}}$ (refer also to the left panel of Fig.~\ref{plotspectra} in the main text, where the evolved quasiparticle is around $k_x = 0$).

It is known that the linear perturbation theory holds only far from population inversion, which means here within the limit $t-t_0 \ll \frac{\pi}{2} \frac{\hbar}{\Delta} \ll \frac{\pi}{2} \frac{\hbar}{V_0}$. In such a limit, from Eq.~\eqref{eqlin} we obtain:
\begin{equation}
\frac{|\langle\Psi (t)|\Omega \rangle|^2 - |\langle \Psi |\Omega \rangle|^2}{\langle\Omega|\Psi\rangle} = - 2 (t-t_0) V_0 \, {\rm Im} \,s[\hat{o}_{\bf k \to 0}] + O( (t-t_0)^2 ) \, ,
\end{equation}
namely
\begin{equation}\label{eq:approximated_imaginary_part}
{\rm Im} \,s[\hat{o}_{\bf k \to 0}] \approx -\frac{1}{2 (t-t_0) V_0}\frac{|\langle\Psi (t)|\Omega \rangle|^2 - |\langle \Psi |\Omega \rangle|^2}{\langle\Omega|\Psi\rangle} \, ,
\end{equation}
that holds up to terms linear in $t-t_0$. Interestingly, Eq.~\eqref{eq:approximated_imaginary_part} can be obtained also by considering the full Rabi evolution
\begin{equation}
\ket{\Psi(t)} = \cos \Big(\frac{V_0}{\hbar} (t-t_0) \Big) \ket{\Psi} + i \,   \sin \Big(\frac{V_0}{\hbar} (t-t_0) \Big) \ket{\Upsilon_{{\bf k} \to 0}} \, ,
\end{equation}
and expressing the left-hand-side of Eq.~\eqref{eqlin} up to the linear order in $t-t_0$.

The formulation above can be equally repeated in coordinate space, depending on ${\bf r}$ and ${\bf r}^{\prime}$, as in Eq.~\eqref{matper}. Particularly relevant is the case, reported in the right panel of Fig.~\ref{plotspectra}, where open boundary conditions are assumed along a direction. In such a case, the term in Eq.~\eqref{pert} makes an edge mode below the half-filling energy to translate into a mode above it, changing its momentum. However, notice that the same perturbation does not induce a net edge current.

\end{document}